# Direct observation of Born-Oppenheimer approximation breakdown in carbon nanotubes

*Adam W. Bushmaker[‡], Vikram V. Deshpande[§], Scott Hsieh[§], Marc W. Bockrath[§], Stephen B. Cronin[‡]\**

Receipt date: 11/25/2008

[‡] University of Southern California, Department of Electrical Engineering - Electrophysics

Los Angeles, CA 90089

[§] California Institute of Technology, Applied Physics.

Pasadena, CA 91125

\*Corresponding Author: Stephen Cronin

Department of Electrical Engineering, University of Southern California

Powell Hall of Engineering PHE 624, Los Angeles, CA 90089-0271

Phone: 213-740-8787

Email: scronin@usc.edu

**Abstract:**

Raman spectra and electrical conductance of individual, pristine, suspended, metallic single-walled carbon nanotubes are measured under applied gate potentials. The $G_-$ band is observed to downshift with small applied gate voltages, with the minima occurring at $E_F = \pm\frac{1}{2}E_{phonon}$, contrary to adiabatic predictions. A subsequent upshift in the Raman frequency at higher gate voltages results in a "W"-shaped Raman shift profile that agrees well with a non-adiabatic phonon renormalization model. This behavior constitutes the first experimental confirmation of the theoretically predicted breakdown of the Born-Oppenheimer approximation in individual single walled carbon nanotubes.

The Born-Oppenheimer (BO) or adiabatic approximation is widely used to simplify the very complex many-body problem of electrons in solids and molecules[1], assuming that electrons equilibrate much faster than the atomic motion of the ionic cores. Without this approximation, most molecular and solid state problems become difficult or impossible to solve analytically. Although the BO approximation is valid in most materials and molecular systems, there are a few situations in which it does not hold, including some low atomic weight compounds[2-4], intercalated graphite[5], and graphene[6]. Clean, defect-free single-walled carbon nanotubes (SWNTs) are systems which can be used to verify fundamental phenomena such as Wigner crystallization[7] and spin-orbit coupling[8], and are ideal candidates for testing fundamental physical predictions. In nanotubes, the BO approximation is expected to break down because of the relatively short vibrational period of the longitudinal optical (LO) phonon and the relatively long electronic relaxation time[9, 10]. This breakdown has been observed in semiconducting nanotube mats[9], however, inhomogeneities broaden effects in such systems.

The breakdown of the BO approximation can be observed directly in an individual nanotube by studying the LO phonon $G_-$ Raman feature of metallic SWNTs (m-SWNTs), which is fundamentally different than that of their semiconducting counterparts[11] (sc-SWNT). The $G_-$ band is broadened and downshifted (reduced in frequency), an effect arising from coupling to a continuum of electronic states[9, 10, 12-18]. In other words, the LO phonon mode is damped by the free electrons near the Fermi energy[19, 20]. This coupling is a Kohn anomaly (KA) and has also been referred to as a weakened Peierl's-like mechanism. The $G_-$ band Raman feature in m-SWNTs is particularly interesting under applied gate voltages ($V_g$) because of the ability to effectively turn off the Kohn anomaly by shifting the Fermi energy ($E_F$). As this happens, the LO phonon frequency upshifts, due to reduced phonon softening of the extinguished Kohn anomaly. This effect has been observed by many groups[13, 15, 16], and generally agrees with phonon renormalization theory quite well[9, 17, 18, 21]. Selected Raman $G$ band spectra from the nanotubes in this study are shown in Figure 1. Note the complete absence of the defect-related $D$ band, which gives testament to the pristine nature of the nanotubes used in this study.

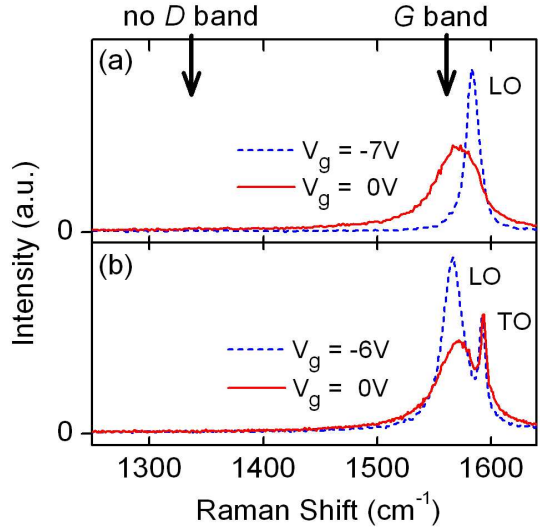

**Figure 1: Sample *G* band spectra from two suspended m-SWNTs showing the Kohn anomaly deactivation.** The spectra in 1(a) correspond to the data plotted in Figure 3, while those in 1(b) correspond to Figures 4 and 5. Note the complete absence of the defect-mediated *D* band.

A striking difference between the predictions of the adiabatic and non-adiabatic models occurs in the gate voltage response of the LO phonon when $E_F$ is near the Dirac point. Phonon energy renormalization calculations done within the adiabatic approximation predict a *monotonic upshift* of the phonon frequency with increasing $|E_F|$, with the minimum frequency at $E_F = 0$ eV (see ref. [10]). In calculations which relax the adiabatic approximation, there is an initial frequency *downshift* of the Γ-LO phonon with increasing $|E_F|$, followed by an upshift after the Fermi energy has passed $\pm\tfrac{1}{2}E_{phonon}$, forming a "W"-shaped gate voltage profile. The minimum frequency for these calculations occurs at $E_F = \pm\tfrac{1}{2}E_{phonon}$[10]. This was recently observed in graphene at cryogenic temperatures[22], however until now there has been no experimental observation of this clear signature of the influence of the BO approximation in isolated SWNTs, most likely because of sample inhomogeneities and defect-related electron relaxation. We report the observation of this initial downshift at room temperature in pristine, isolated m-SWNTs, followed by a subsequent upshift, consistent with theoretical predictions reported previously[10, 18]. These results directly confirm the breakdown of the Born-Oppenheimer approximation, indicating the intrinsic non-adiabatic nature of the electron-phonon coupling in this system.

The *G-* band Raman peak is often observed to be asymmetric, consistent with a Breit-Wigner-Fano (BWF) lineshape, given by $I_{BWF}(\omega) = I_0 \frac{(\gamma + (\omega-\omega_0)\frac{1}{q})^2}{\gamma^2 + (\omega-\omega_0)^2}$, where $\gamma$ is the linewidth $\omega_o$ is the center frequency, and $q$ is the Fano factor (negative in m-SWNTs). This asymmetry is due to photon coupling to a discrete phonon state and to a continuum of electronic states[12]. The temperature[23] and gate voltage[17, 24] dependences of this parameter have been previously reported. Here, we present the gate voltage dependence of *-1/q* and electrical conductance measured simultaneously.

In this work, samples are fabricated using chemical vapor deposition on Pt electrodes with predefined catalyst beds, as reported previously[7, 25, 26]. The resulting devices are suspended single-walled nanotubes with a trench depth of 300nm and widths of 2-5 µm, illustrated in Figure 2. The samples in this study were grown using ethanol[27] or methane as the carbon feedstock. No additional processing was performed on devices after the nanotube growth, except for an oxygen bake to rid the devices of amorphous carbon. The devices are pre-screened by examination of the electrical characteristics, such that all nanotubes in this study are highly defect-free, pristine, individual SWNT devices. Raman spectra were collected from the center of the SWNTs with a Renishaw *InVia* spectrometer using 532nm, 633nm, or 785nm lasers focused to a diffraction limited spot through a 100X, high numerical aperture objective lens.

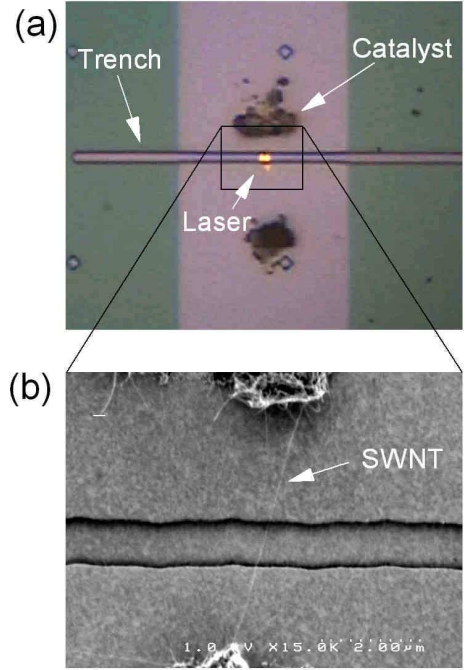

**Figure 2: Device Geometry:** (a) Optical microscope image showing SWNT device with focused laser spot and (b) close-up SEM image of the device, showing the trench spanned by a suspended SWNT.

Figure 3 shows the Raman frequency, linewidth, $-1/q$, and electrical conductance data plotted as a function of applied gate voltage ($V_g$) and Fermi energy ($E_F$), as determined from the gate coupling factor. The laser wavelength used was 532nm at a power of 350μW. The frequency of the LO phonon initially downshifts as $|E_F|$ is moved away from zero. As $|E_F|$ is increased beyond $E_{ph}/2$, the Raman frequency (Figure 3(a)) begins to upshift. The Raman linewidth (full-width half-max) of the LO $G_-$ band is also plotted versus $V_g$ and $E_F$ in Figure 3(b), and exhibits a strong narrowing as the Kohn anomaly is shut off with increasing $|E_F|$, dropping from over 50 cm$^{-1}$ to just over 10 cm$^{-1}$. Plotted along with the Raman data are the adiabatic (dashed) and non-adiabatic (solid) phonon renormalization models presented by Caudal *et al*[10] and described below. Finally, $-1/q$ exhibits a strong decrease towards zero with increasing $|E_F|$, while the conductance shows a sharp dip near $E_F = 0$, typical of the quasi-metallic nanotubes measured in this study. The solid line in Figure 3c represents a conductivity model based on Boltzmann transport and the Landauer model (Equation (3)) [26, 28-33].

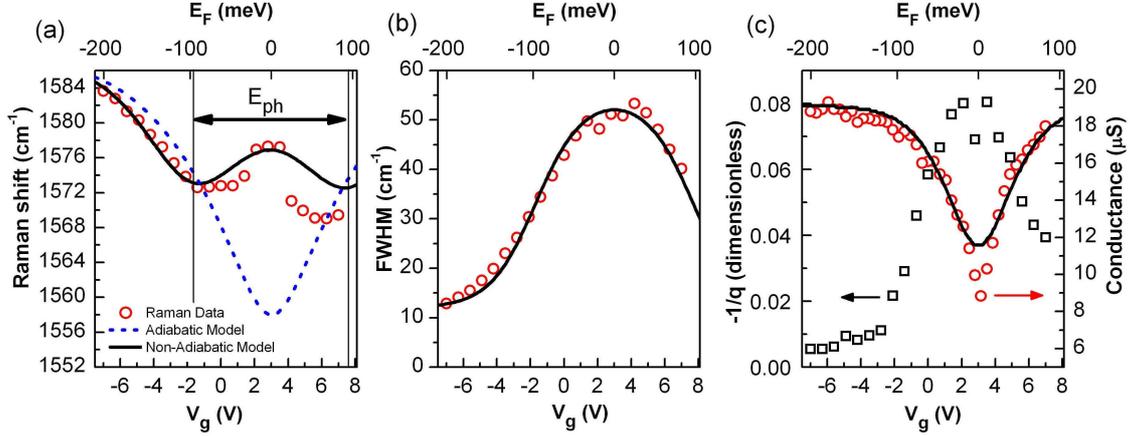

**Figure 3: Raman spectral data from a single walled carbon nanotube.** LO (a) shift (filled circles), (b) linewidth (FWHM, filled circles) and (c) -1/$q$ (open squares), as well as the electrical conductivity (filled circles) are plotted versus gate voltage ($V_g$). The Fermi energy ($E_F$) is also indicated on the top $x$-axis. The lines in (a) and (b) show the results of the adiabatic (dashed) and non-adiabatic (solid) phonon renormalization models discussed below (Equations (1-2)), while the solid line in (c) represents the Boltzmann-Landauer transport model (Equation (3)).

Figures 4 and 5 show the gate voltage dependence of the Raman and conductance data of another suspended nanotube, taken with a 633nm laser at a power of 1mW. The upshift at large |$E_F$| is experimentally difficult to observe, because of the relatively weak gate coupling, and the fact that suspended nanotubes are eventually destroyed at high gate voltages. A strong initial frequency downshift was observed in all 6 nanotubes of this study. The values reported here for -1/$q$ are smaller than the values reported elsewhere[12, 17, 23, 24], where -1/$q$ ~ 0.2-0.4. An interesting trend is that the gate voltage dependence of -1/$q$ was noticed to change proportionally to (FWHM-$\gamma_0$)$^2$, where $\gamma_0$ is the intrinsic linewidth of 10-15 cm$^{-1}$.

The -1/$q$ value for this nanotube is substantially smaller than that of the previous nanotube (shown in Figure 3(c)), and was observed to decrease over the course of the sample's lifetime (several months in air and at high temperatures during testing) from $-1/q = 0.07$ at $E_F = 0$ to $-1/q = 0.02$ at $E_F = 0$. The maximum FWHM for this nanotube also decreased, however, only by 2 cm$^{-1}$ from 44.4 cm$^{-1}$ to 42.5 cm$^{-1}$. From the absence of a $D$ band Raman signal throughout the experiment, we can infer that few defects were introduced into the SWNT over this time. However the large variation in $-1/q$ indicates that it is extremely sensitive to aging and environmental conditions; in fact far more sensitive than the FWHM or the $D$ band intensity. These experimental results highlight the need for a quantitative model describing the behavior of the Fano parameter $-1/q$ in response to gate voltages and environment changes.

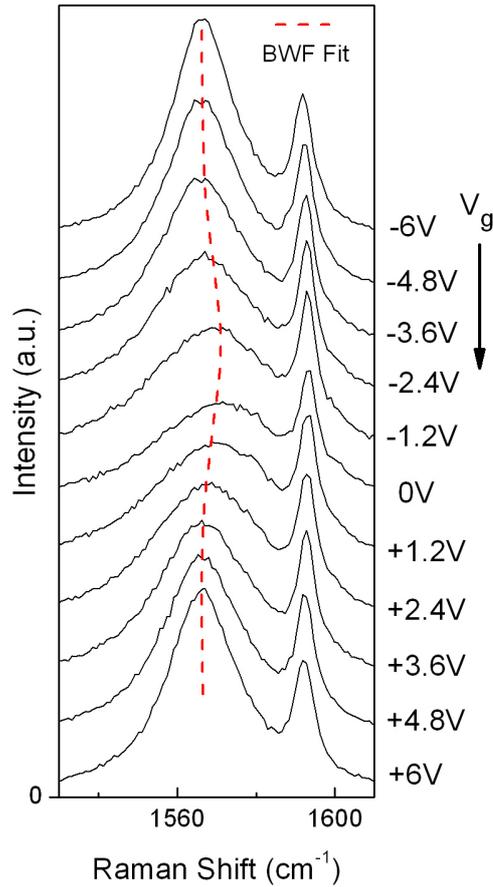

**Figure 4: Gate voltage evolution of the Raman spectra from a second device.** The gate voltage dependence of the parameters fit to these spectra are presented in Figure 5(a-c).

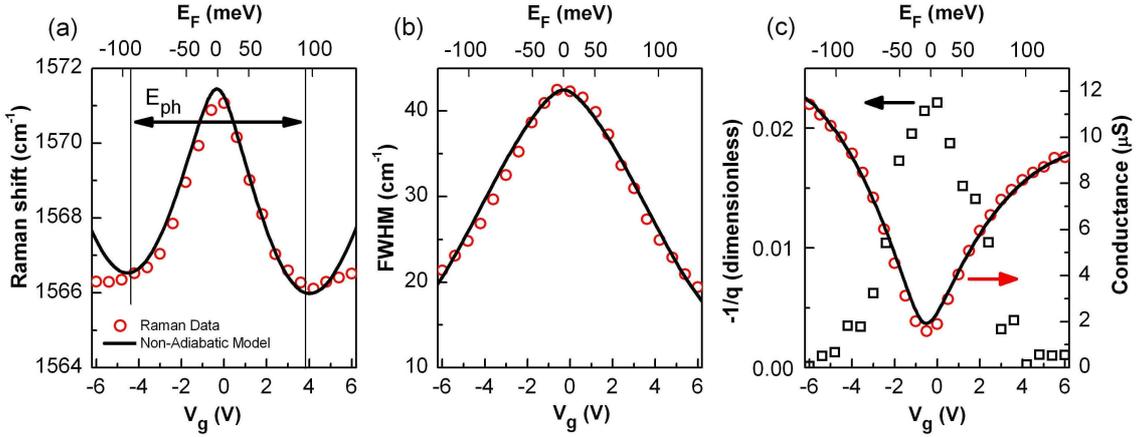

**Figure 5: Raman spectral data from a second SWNT device.** (Spectra in Figure 4) Raman LO (a) shift (filled circles), (b) linewidth (FWHM, filled circles) and (c) -1/q (open squares), as well as the electrical conductivity (filled circles) are plotted versus gate voltage ($V_g$) and Fermi energy ($E_F$).

The non-adiabatic phonon renormalization model used in Figures 3 and 5 was confirmed by density functional theory (DFT) calculations[10], and is outlined below. The equation describing the frequency of the Γ-point optical phonon, $\omega_\Gamma$, is given by

$$\omega_\Gamma = \sqrt{\frac{D_\Gamma^0}{M} + \frac{D_\Gamma^{KA}}{M}}, \quad (1)$$

where $M$ is the mass of the carbon atom and $D_\Gamma^0$ is the intrinsic (no Kohn anomaly) dynamical matrix. The equation for the non-analytical electron-phonon coupling contribution to the dynamical matrix, $D_\Gamma^{KA}$, is given by

$$D_\Gamma^{KA} = \frac{a_0^2 2\sqrt{3}}{\pi^2 d_t}\langle D_\Gamma^2\rangle_F \int_{-\bar{k}}^{\bar{k}}\left\{\frac{f[\varepsilon_v(k')]-f[\varepsilon_c(k')]}{\varepsilon_v(k')-\varepsilon_c(k')+E_{ph}^{LO,\Gamma}+i\delta}+\frac{f[\varepsilon_c(k')]-f[\varepsilon_v(k')]}{\varepsilon_c(k')-\varepsilon_v(k')+E_{ph}^{LO,\Gamma}+i\delta}\right\}dk', \quad (2)$$

where $a_0$ is the graphene lattice constant, $E_{ph}^{LO,\Gamma}$ is the phonon energy, $d_t$ is the nanotube diameter, $\langle D_\Gamma^2 \rangle_F$ is the electron-phonon coupling constant, $\bar{k}$ is a small integration limit, $f$ is the Fermi function, $\varepsilon_v(k')$ and $\varepsilon_c(k')$ are the hyperbolic valence and conduction band dispersion relationships, respectively, and $\delta$ is the electronic lifetime broadening coefficient, found to be 0.9 meV and 0.8 meV for the nanotubes in Figures 3 and 5, respectively. The adiabatic case is approximated simply by setting $E_{ph}^{LO,\Gamma} = 0$. The FWHM is given by the imaginary part of the non-adiabatic dynamical matrix plus an intrinsic linewidth of 10-15 cm$^{-1}$. It is possible that this effect of non-adiabatic phonon hardening near $E_F = 0$ has not been observed until now because the substrate interaction increases the electronic scattering rate by defect scattering. This would make the Born-Oppenheimer approximation valid, by eliminating the non-adiabatic phonon hardening near $E_F = 0$. Also, the effect would be difficult to observe bulk samples due to inhomogeneities in the Fermi level[9].

From the diameter of the nanotube measured (obtained from the RBM frequency), the electron-LO phonon coupling constant $\langle D_\Gamma^2 \rangle_F$ can be found directly by fitting the data. The nanotube in Figure 3 exhibited an RBM in its spectra at 115.6 cm$^{-1}$, corresponding to a diameter of 1.97 nm according to the relation[34] $d_t = 227/\omega_{RBM}$, giving a value of $\langle D_\Gamma^2 \rangle_F = 46$ (eV/Å)$^2$. Another nanotube (not shown) exhibited an RBM at 174.0 cm$^{-1}$, corresponding to a diameter of 1.31 nm and $\langle D_\Gamma^2 \rangle_F = 52$ (eV/Å)$^2$.

The electrical data in Figures 3(c) and 5(c) are fit to the Landauer model using the Boltzmann equation in the constant relaxation time approximation[26, 28-32]. The resistance of the nanotube can be found by taking a sum of the phonon scattering resistance and the quantum resistance,

$$R(V,T) = R_{scatt} + R_{quantum}, \tag{3}$$

where each contribution is found by summing over the density of states near the Fermi energy, following Biercuk and McEuen[32]. The Fermi energy is calculated numerically as a function of gate voltage using a geometric gate capacitance $C$, the Fermi function, and a hyperbolic density of states model[33], according to the equation $E_F + \frac{Q(E_F)}{C} = eV_g$, where $Q$ is the charge induced on the nanotube. This equation includes the effect of the mini band-gap (where the density of states is zero), which creates a non-linear $V_g$-$E_F$ relationship. Inclusion of this non-linear relationship in the model is key to fitting the data properly. The mean free path of electrons scattering by acoustic phonons, $\lambda_{ac}$, was taken to be 2 μm, in accordance with previous publications[30]. The data was fit to the frequency, width, and conductivity models self-consistently, with the gate capacitance, $\langle D_\Gamma^2 \rangle_F$, $d_t$, $\delta$, contact transmission coefficients, and mini-bandgap as fitting parameters. The bandgaps for the SWNTs in Figures 3 and 5 were found to be 42 meV and 120meV, respectively. For our single, pristine SWNTs, the model can be seen to fit the data reasonably well with a gate capacitance of 1.5-1.8 pF/m.

In conclusion, we report Raman spectra of isolated, suspended metallic SWNTs (m-SWNTs) observed with applied gate voltages. The LO phonon Raman band ($G_-$) is observed to initially downshift with applied gate voltage, then subsequently upshift for $|E_F| > ½ E_{ph}^{LO,\Gamma}$. This behavior is attributed to the non-adiabaticity of the Γ-point Kohn anomaly in m-SWNTs, and constitutes the first experimental confirmation of the predicted breakdown of the Born-Oppenheimer approximation in individual SWNTs. The Raman data agree quantitatively with a non-adiabatic model using time-dependent perturbation theory, while the electron transport data are fit using the Landauer model and the Boltzmann equation within the constant relaxation time approximation. The results showcase the use of pristine, defect-free nanotubes as model systems for studying fundamental phenomena.

This research was supported in part by DOE Award No. DE-FG02-07ER46376 and the National Science Foundation Graduate Research Fellowship Program. We would like to thank F. Mauri and M. Saitta for discussions concerning the implementation of the phonon energy renormalization model.

1. Born, M.; Oppenheimer, R. *Annalen Der Physik* **1927,** 84, (20), 0457-0484.
2. Vidal-Valat, G.; Vidal, J. P.; Kurki-Suonio, K.; Kurki-Suonio, R., Evidence on the breakdown of the Born-Oppenheimer approximation in the charge density of crystalline 7LiH/D. In *Acta Crystallographica Section A*, 1992; Vol. 48, pp 46-60.
3. Cappelluti, E.; Pietronero, L. *SMEC 2005, Study of matter under extreme conditions* **2006,** 67, (9-10), 1941-1947.
4. Che, L.; Ren, Z.; Wang, X.; Dong, W.; Dai, D.; Wang, X.; Zhang, D. H.; Yang, X.; Sheng, L.; Li, G.; Werner, H.-J.; Lique, F.; Alexander, M. H. *Science* **2007,** 317, (5841), 1061-1064.
5. Saitta, A. M.; Lazzeri, M.; Calandra, M.; Mauri, F. *Phys. Rev. Lett.* **2008,** 100, (22), 226401-4.
6. Pisana, S.; Lazzeri, M.; Casiraghi, C.; Novoselov, K. S.; Geim, A. K.; Ferrari, A. C.; Mauri, F. *Nature Materials* **2007,** 6, 198–201.
7. Deshpande, V. V.; Bockrath, M. *Nat Phys* **2008,** 4, (4), 314-318.
8. Kuemmeth, F.; Ilani, S.; Ralph, D. C.; McEuen, P. L. *Nature* **2008,** 452, (7186), 448-452.
9. Das, A.; Sood, A. K.; Govindaraj, A.; Saitta, A. M.; Lazzeri, M.; Mauri, F.; Rao, C. N. R. *Phys. Rev. Lett.* **2007,** 99, (13), 136803.
10. Caudal, N.; Saitta, A. M.; Lazzeri, M.; Mauri, F. *Phys. Rev. B* **2007,** 75, (11), 115423-11.
11. Dresselhaus, M. S.; Dresselhaus, G.; Jorio, A.; Souza Filho, A. G.; Saito, R. *Carbon* **2002,** 40, (12), 2043-2061.
12. Brown, S. D. M.; Jorio, A.; Corio, P.; Dresselhaus, M. S.; Dresselhaus, G.; Saito, R.; Kneipp, K. *Phys. Rev. B* **2001,** 63, (15), 155414.
13. Rafailov, P. M.; Maultzsch, J.; Thomsen, C.; Kataura, H. *Phys. Rev. B* **2005,** 72, (4), 45411.
14. Wu, Y.; Maultzsch, J.; Knoesel, E.; Chandra, B.; Huang, M.; Sfeir, M. Y.; Brus, L. E.; Hone, J.; Heinz, T. F. *Phys. Rev. Lett.* **2007,** 99, (2), 027402-4.

15. Kavan, L.; Rapta, P.; Dunsch, L.; Bronikowski, M. J.; Willis, P.; Smalley, R. E. *J. Phys. Chem. B* **2001,** 105, (44), 10764-10771.
16. Cronin, S. B.; Barnett, R.; Tinkam, M.; Chou, S. G.; Rabin, O.; Dresselhaus, M. S.; Swan, A. K.; Ünlü, M. S.; Goldber, B. B. *Appl. Phys. Lett.* **2004,** 84, (12), 2052.
17. Farhat, H.; Son, H.; Samsonidze, G. G.; Reich, S.; Dresselhaus, M. S.; Kong, J. *Phys. Rev. Lett.* **2007,** 99, (14), 145506-4.
18. Tsang, J. C.; Freitag, M.; Perebeinos, V.; Liu, J.; Avouris, P. *Nature Nanotechnology* **2007,** 2, (11), 725-730.
19. Dubay, O.; Kresse, G.; Kuzmany, H. *Phys. Rev. Lett.* **2002,** 88, (23), 235506.
20. Piscanec, S.; Lazzeri, M.; Robertson, J.; Ferrari, A. C.; Mauri, F. *Phys. Rev. B* **2007,** 75, (3), 35427.
21. Sasaki, K.-i.; Saito, R.; Dresselhaus, G.; Dresselhaus, M. S.; Farhat, H.; Kong, J. *Phys. Rev. B* **2008,** 77, (24), 245441-8.
22. Yan, J.; Henriksen, E. A.; Kim, P.; Pinczuk, A. *Arxiv preprint arXiv:0712.3879* **2007**.
23. Uchida, T.; Tachibana, M.; Kurita, S.; Kojima, K. *Chem. Phys. Lett.* **2004,** 400, (4-6), 341-346.
24. Nguyen, K. T.; Gaur, A.; Shim, M. *Phys. Rev. Lett.* **2007,** 98, (14), 145504-4.
25. Bushmaker, A. W.; Deshpande, V. V.; Bockrath, M. W.; Cronin, S. B. *Nano Lett.* **2007**.
26. Pop, E.; Mann, D.; Cao, J.; Wang, Q.; Goodson, K.; Dai, H. *Phys. Rev. Lett.* **2005,** 95, (15), 155505.
27. Huang, L. M.; White, B.; Sfeir, M. Y.; Huang, M. Y.; Huang, H. X.; Wind, S.; Hone, J.; O'Brien, S. *J. Phys. Chem. B* **2006,** 110, (23), 11103-11109.
28. Seri, T.; Ando, T. *Journal of the Physical Society of Japan* **1997,** 66, (1), 169-173.
29. Ando, T. *Journal of the Physical Society of Japan* **2005,** 74, (3), 777-817.
30. Park, J. Y.; Rosenblatt, S.; Yaish, Y.; Sazonova, V.; Ustunel, H.; Braig, S.; Arias, T. A.; Brouwer, P. W.; McEuen, P. L. *Nano Lett.* **2003,** 4, (3), 517.
31. Yao, Z.; Kane, C. L.; Dekker, C. *Phys. Rev. Lett.* **2000,** 84, (13), 2941.
32. Biercuk, M. J.; Ilani, S.; Marcus, C. M.; McEuen, P. L. *TOPICS IN APPLIED PHYSICS* **2008,** 111, 455.
33. Bockrath, M.; Cobden, D. H.; McEuen, P. L.; Chopra, N. G.; Zettl, A.; Thess, A.; Smalley, R. E. *Science* **1997,** 275, (5308), 1922-1925.
34. Araujo, P. T.; Maciel, I. O.; Pesce, P. B. C.; Pimenta, M. A.; Doorn, S. K.; Qian, H.; Hartschuh, A.; Steiner, M.; Grigorian, L.; Hata, K.; Jorio, A. *Phys. Rev. B* **2008,** 77, (24), 241403-4.